\documentclass[aps,prl,twocolumn,
]{revtex4-1}

\usepackage{amssymb}
\usepackage{amsmath}
\usepackage{float}
\usepackage{graphicx}				
\usepackage[caption=false]{subfig}
\usepackage{mathrsfs}
\usepackage{multirow}
\usepackage{color}
\usepackage{longtable}

\begin{document}

\title{Laser stabilization using saturated absorption in a cavity QED system} 
\author{D. A. Tieri$^1$}
\author{J. Cooper$^1$}
\author{Bjarke T. R. Christensen$^2$ }
\author{J. W. Thomsen$^2$}
\author{M. J. Holland$^1$}
\affiliation{$^1$JILA, National Institute of Standards and Technology and University of Colorado, Boulder, CO 80309-0440, USA \\
$^2$Niels Bohr Institute, University of Copenhagen; Blegdamsvej 17, 2100 Copenhagen, Denmark}
\date{\today}

\begin{abstract}
We consider the phase stability of a local oscillator (or laser) locked to a cavity QED system comprised of atoms with an ultra-narrow optical transition. The atoms are cooled to millikelvin temperatures and then released into the optical cavity. Although the atomic motion introduces Doppler broadening, the standing wave nature of the cavity causes saturated absorption features to appear, which are much narrower than the Doppler width. These features can be used to achieve an extremely high degree of phase stabilization, competitive with the current state-of-the-art. Furthermore, the inhomogeneity introduced by finite atomic velocities can cause optical bistability to disappear, resulting in no regions of dynamic instability and thus enabling a new regime accessible to experiments where optimum stabilization may be achieved. 
\end{abstract}

\maketitle

\section{Introduction}
Today's ultra-precise and accurate atomic clocks continue to make important contributions to fundamental physics as well as applied technology. Atomic clocks have imposed significant constraints on the drift of fundamental constants \cite{LeTargat,Huntemann,Godun}, may have the potential to enhance the sensitivity of gravitational wave detectors, and have provided ultimate tests of the general theory of relativity \cite{Schiller}.  With the current stability of optical clocks at the 1$\times10^{-18}$ level, there are prospects for applying atomic clocks for the detailed mapping of the Earth's gravity field \cite{Guena,Bondarescu01102012}.

A highly stabilized laser is an integral component of high precision measurements, such as optical atomic clocks and precision spectroscopy. Current technology for achieving highly phase stable laser sources relies on locking a laser to a high-Q reference ULE glass cavity \cite{drever1983laser, young1999visible, jiang2011making}. The phase stability of this method is currently limited by the thermal noise induced in the mirrors, spacers, and coatings of the reference cavity \cite{numata2004thermal}, but has been significantly reduced over the past few years with new engineered materials \cite{Kessler, Martin}.
 
As an alternative approach to overcoming the thermal noise problem, it was recently proposed \cite{PhysRevA.84.063813} to lock the laser to the saturated resonance feature exhibited by a collection of atoms with an ultra-narrow electronic transition trapped in an optical cavity. Here, the atoms were assumed to be trapped in an optical lattice inside the cavity. Due to the narrow atomic line, such a system would typically operate in the parameter region corresponding to the bad cavity limit of cavity QED. There, the atomic linewidth is significantly narrower than the cavity linewidth. In contrast to the reference cavity stabilization method described above, the cavity QED method offers a distinct advantage since no drift compensation is needed.

The cavity QED system exhibits optical bistability in the intracavity intensity \cite{PhysRevA.18.1129, 0305-4470-13-2-034} where several solutions exist for the steady-state intracavity field. Working at an input intensity in the region where bistability is present would in principle allow the greatest degree of stabilization \cite{PhysRevA.84.063813}. However, it is not practical to work in the bistable region since quantum and classical fluctuations between the semi-classical eigenmodes cause the system to be dynamically unstable. Therefore, one is restricted to working at input intensities above the bistability, where the achievable stabilization is orders of magnitude worse. Still, it was shown \cite{PhysRevA.84.063813} that phase stability corresponding to the sub-mHz level should be achievable.

\begin{figure}[t]
\begin{center}
\includegraphics[width=8.3cm,height=3.8cm] {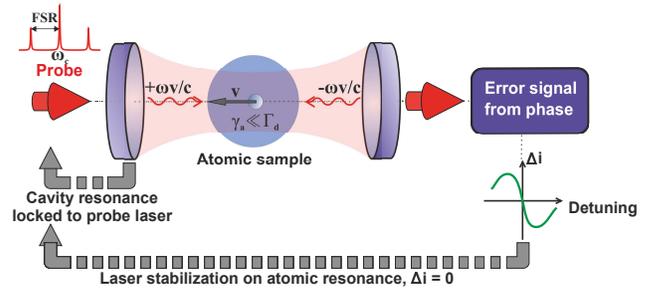}
\end{center}
\caption{(color online). Schematic of the cavity QED experiment with a thermal sample of atoms with Doppler width $\Gamma_d$, each of which have a narrow optical transition of width $\gamma_a$ $<<$ $\Gamma_d$. The atoms are probed with a carrier with frequency $\omega_{c}$, and two sidebands located at $\omega_{c}\pm$FSR, where FSR is the free spectral range of the cavity. The carrier frequency $\omega_{c}$, which is close to the atomic frequency $\omega_{a}$, is locked on the cavity mode frequency $\omega$, while the sidebands are assumed far off resonance, typically $\sim10^4$ atomic linewidths. By demodulating the light transmitted through the cavity at the FSR, detection of the non-linear phase response of the transmitted light is achieved. This phase response results in a photocurrent that serves as a frequency discriminating error signal that stabilizes the laser frequency through the requirement $\Delta i=0$.}
\label{f1}
\end{figure}

Recently \cite{JanPRL}, an experimental effort to demonstrate the cavity QED system was made by probing the $\left| ^1S_0\right> - \left| ^3P_1\right>$ intercombination line of $^{88}$Sr atoms (i.e. $\gamma_a/2\pi$ = 7.6 kHz) inside an optical cavity. There, however, the atoms were not trapped in an optical lattice, but loaded into the center of the cavity using a MOT, which was then turned off during probing. Typical MOT temperatures correspond to a few millikelvin which is equivalent to a Doppler width of several MHz. Considering the narrow 7.6 kHz line of the optical transition, this implies that motional effects will be important.

In this paper, we extend the many-atom cavity QED theory of \cite{PhysRevA.84.063813} to include atomic motion, and study its effect on the stabilization precision. In spite of the large Doppler effect, the standing-wave nature of the cavity field induces sharp saturated absorption and dispersion features to appear in the considered observables. These features are nestled in the center of the overall Doppler broadened features \cite{Stenholm1969, Freed, Stenholm, levenson2012introduction}.  The stabilization that is achievable by utilizing these sharp features is impeded by multi-photon scattering processes that occur when an atom's velocity matches one of its Doppleron resonances \cite{PhysRevA.46.479,Tallet:94}. The dependence of the stabilization on the number of atoms and the temperature due to the Dopplerons is discussed. We demonstrate that the motion of the atoms causes the bistability region to disappear, so that no restrictions on input power are necessary to avoid the dynamic instability that would otherwise result.

\section{Model}
We model our system as a collection of $N$ two-level atoms inside a single
mode optical cavity using the quantum Born-Markov master equation to describe
the open quantum system,
\begin{equation}
  \frac{d}{dt} \hat{\rho} = \frac{1}{i \hbar} \left[ \hat{H}, \hat{\rho} \right] + \hat{\mathcal{L}}\left[ \hat{\rho} \right],
\label{ME1}
\end{equation}
where,
\begin{equation}
  \hat{H} = \frac{\hbar \Delta}{2} \sum_{j=1}^{N} \hat{\sigma}^{z}_{j}
  + \hbar \eta \left( \hat{a}^{\dagger} + \hat{a}  \right)
  + \hbar  \sum_{j=1}^{N} g_j(t) \left( \hat{a}^{\dagger} \hat{\sigma}^{-}_{j}
    + \hat{\sigma}^{+}_{j} \hat{a} \right)\,,
\end{equation}
and $\hat{\mathcal{L}}\left[ \hat{\rho} \right]$ denotes the Liouvillian.

The Hamiltonian $H$ describes the coherent evolution of the coupled atom cavity system in an interaction picture which rotates at the frequency of the cavity, and $\Delta$ is the atom-cavity detuning. The Pauli spin matrices for the atoms are $\hat{\sigma}_j^{+}$ ,$\hat{\sigma}_j^{-}$ and $\hat{\sigma}_j^{z}$, and $\hat{a}$ is the annihilation operator of the cavity mode. Furthermore, $\eta=\sqrt{(\kappa P_{in})/(\hbar \omega)}$ is the classical drive amplitude, where $\kappa$ is the decay rate of the cavity, $P_{in}$ is the input power, and $\omega$ is the frequency of the cavity mode. The atom-cavity coupling rate is $g_j(t)= g_0  \cos(\delta_jt)$, where $g_0$ is the maximum coupling amplitude, and $\delta_j=kv_j$ is the Doppler shift in terms of the velocity $v_j$ of the $j$th atom, and wave number $k$ of the light.

The incoherent evolution describes the various forms of dissipation in this system and is described by the Liouvillian $\hat{\mathcal{L}}\left[ \hat{\rho} \right]$,
\begin{eqnarray}
  \hat{\mathcal{L}}\left[ \hat{\rho} \right] &=&
  -\frac{\kappa}{2} \left\{ \hat{a}^{\dagger} \hat{a} \hat{\rho}
    + \hat{\rho}  \hat{a}^{\dagger} \hat{a}
    - 2 \hat{a} \hat{\rho} \hat{a}^{\dagger} \right\}
  \nonumber
  \\
  &&-\frac{\gamma}{2} \sum_{j=1}^N \left\{ \hat{\sigma}_{j}^{+}
    \hat{\sigma}_{j}^{-} \hat{\rho}
    + \hat{\rho} \hat{\sigma}_{j}^{+} \hat{\sigma}_{j}^{-}
    - 2  \hat{\sigma}_{j}^{-} \hat{\rho} \hat{\sigma}_{j}^{+}
  \right\}
  \nonumber
  \\
  &&+\frac{\gamma_p}{2} \sum_{j=1}^N \left\{ \  \hat{\sigma}_{j}^{z}
    \hat{\rho}  \hat{\sigma}_{j}^{z} -  \hat{\rho}
  \right\},
\end{eqnarray}
where $\hat{\rho}$ is the system's density matrix, $\gamma$ is the spontaneous emission rate for the atoms, and $\gamma_p$ is the total decay rate of the atomic dipole.

We derive Langevin equations corresponding to Eq.~(\ref{ME1}). Assuming that the classical drive $\eta$ is sufficiently strong, a mean-field description provides an accurate representation \cite{PhysRevA.56.3262}. We define the mean values for the field $\alpha = - i \left< \hat{a} \right>$, and for the atoms,
$\sigma^{-}_{j}= \left< \hat{\sigma}^{-}_{j} \right>, \sigma^{+}_{j}=
\left< \hat{\sigma}^{+}_{j} \right> , \sigma^{z}_{j}= \left<
  \hat{\sigma}^{+}_{j} \right>$, which evolve according to the semiclassical evolution,
\begin{eqnarray}
  \dot{\alpha}&=&-\kappa \alpha + \eta +
  \sum_{j=1}^{N} g_j(t) \sigma^{-}_{j},
\label{a1}
\\
\dot{\sigma}^{-}_{j} &=&-\left( \gamma_p
  + i \Delta \right) \sigma^{-}_{j} +  g_j(t) \alpha \sigma^{z}_{j},
\label{sm1}
\\
\dot{\sigma}^{z}_{j} &=& -\gamma \left( \sigma^{z}_{j}
  + 1 \right) - 2  g_j(t) \left( \alpha \sigma^{+}_{j}
  +  \alpha^{*} \sigma^{-}_{j} \right).
\label{sz1}
\end{eqnarray}
In the moving frame of reference of the $j$th atom, the cavity field
appears as a traveling wave, containing two frequencies shifted above
and below the cavity frequency by the Doppler shift $\delta_j$ (refer to Fig.~\ref{f1}).

It is convenient to approximate Eqns.~(\ref{a1}--\ref{sz1}) as a function of the continuous variable $\delta = k v$,
\begin{eqnarray}
  \dot{\alpha}&=&-\kappa \alpha + \eta +
  g_0 \int d\delta P(\delta)  \cos(\delta t) \sigma^{-},
\label{a2}
\\
\dot{\sigma}^{-} &=&-\left( \gamma_p
  + i \Delta \right) \sigma^{-} +  g_0  \cos(\delta t) \alpha \sigma^{z},
\label{sm2}
\\
\dot{\sigma}^{z} &=& -\gamma \left( \sigma^{z}
  + 1 \right) - 2  g_0  \cos(\delta t) \left( \alpha \sigma^{+}
  +  \alpha^{*} \sigma^{-} \right),
\label{sz2}
\end{eqnarray}
where $P(\delta)$ is the Maxwell velocity distribution of width $\delta_0$, which is related to the temperature by the equipartition theorem.

To solve this problem that intrinsically contains a bi-chromatic drive, we proceed in two ways. Our first approach is to numerically integrate Eqns.~(\ref{a2}--\ref{sz2}), partitioning the integral into finite-size velocity bins. The velocity partition must be chosen with care, since the system exhibits Doppleron resonances, which have a strong dependence on the atomic velocity. Specifically, at lower velocity, more resolution in the partition is required.

Our second approach is semi-analytic, and involves a Floquet analysis \cite{PhysRevA.48.3092, Agarwal:91}, in which we expand $\sigma^{-}$, $\sigma^{+}$, and $\sigma^{z}$ in terms of their Fourier components,
 \begin{eqnarray}
\sigma^{-}&=& \sum_l e^{i l \delta t} x_1^{(l)},
\nonumber
\\
\sigma^{+}&=& \sum_l e^{i l \delta t} x_2^{(l)},
\nonumber
\\
\sigma^{z}&=& \sum_l e^{i l \delta t} x_3^{(l)},
\label{FloquetExpansion}
\end{eqnarray}
where $x_1^{(l)}$,$x_2^{(l)}$, and $x_3^{(l)}$ are the amplitudes of the $l$th Fourier component. Upon substitution of Eqns.~(\ref{FloquetExpansion})  into Eqns.~(\ref{a2}--\ref{sz2}), equations for the amplitudes are found:

\begin{eqnarray}
\dot{x}^{(l)}_{1} &=&-\left(i (\Delta+l \delta) + \gamma_p \right) x^{(l)}_{1} +  \frac{g_0\alpha}{2} \left( x^{(l+1)}_3 +  x^{(l-1)}_3 \right),
\label{x1}
\nonumber
\\
\\
\dot{x}^{(l)}_{2} &=&\left(i (\Delta+l \delta) - \gamma_p \right) x^{(l)}_{2} +  \frac{g_0 \alpha^{*}}{2} \left( x^{(l+1)}_3 + x^{(l-1)}_3 \right),
\label{x2}
\nonumber
\\
\\
\dot{x}^{(l)}_{3} &=& -\gamma \delta_{l,0} - \left( i l \delta + \gamma \right) x^{(l)}_{3}
\nonumber
\\
&&- g_0\left( \alpha x^{(l+1)}_2 + \alpha^{*} x^{(l+1)}_1 + \alpha x^{(l-1)}_2 + \alpha^{*} x^{(l-1)}_1 \right).
\label{x3}
\nonumber
\\
\end{eqnarray}

In order to find a steady state solution, we set the time derivatives in Eqns.~(\ref{x1}--\ref{x3}) to zero, and substitute Eqns.~(\ref{x1}--\ref{x2}) into Eq.~(\ref{x3}), yielding,

\begin{equation}
0=\gamma \delta_{l,0} + a_l x_3^{(l)} + d_l x_3^{(l+2)} + b_l x_3^{(l-2)},
\label{x3ss}
\end{equation}
where $\delta_{l,0}$ is a Kronecker delta, and
\begin{equation}
a_l \equiv il\delta +\gamma + \frac{g_0^2 |\alpha|^2}{2}\left(\frac{1}{Q_{l+1}}+\frac{1}{P_{l+1}}+\frac{1}{Q_{l-1}}+\frac{1}{P_{l-1}}\right),
\end{equation}
\begin{equation}
b_l = d_{l-2}\equiv \frac{g_0^2 |\alpha|^2}{2}\left(\frac{1}{Q_{l-1}}+\frac{1}{P_{l-1}}\right),
\end{equation}
where,
\begin{equation}
P_l = i(l\delta+\Delta) +\gamma_p,
\end{equation}
\begin{equation}
Q_l = i(l\delta-\Delta) +\gamma_p.
\end{equation}

For a given $\alpha$, Eq.~(\ref{x3ss}) defines a tridiagonal linear system that can be solved by truncating $l$ at some finite value, and applying the Thomas algorithm for matrix inversion \cite{ASC}. 

Since the atoms have motion, the condition for resonance between an atom and photon is achieved when the atomic frequency and photon frequency are offset by $\delta$. However, higher order multi-photon processes, known as Doppleron resonances, involving $2n+1$ photons where $n$ is an integer, are also possible. The n$th$ order Doppleron resonance corresponds to the terms of order $l/2$ in the Floquet theory \cite{PhysRevA.46.479,Tallet:94}. This correspondence is why only even values of $l$ can couple into Eq.~(\ref{x3ss}).

In steady state, Eq.~(\ref{a2}) simplifies to,
\begin{equation}
\alpha = \frac{\eta}{\kappa} +\frac{g_0^2 N}{2\kappa} \int d\delta P(\delta) \left( x^{(-1)}_1 + x^{(1)}_1  \right).
\label{anm}
\end{equation}
Since $x_1^{-1}$ and $x_1^{1}$ depend on $\alpha$, the self-consistent field amplitude $\alpha$ that solves Eq.~(\ref{anm}) is found numerically by applying Newton's method for root finding \cite{ASC}.

We have seen excellent agreement between the two previously described solution methods, and for the remainder of the paper, focus our attention on the Floquet solution, which most transparently illuminates the underlying physics.

\section{Discussion of steady state solutions}

We first consider the lowest order solution to Eq.~(\ref{anm}), by truncating at $l=0$, which means we have not included higher order Doppleron processes. We have verified that this solution displays the correct physics qualitatively by comparing to higher order solutions that are truncated at increasing values of $l$. We define scaled intracavity and input field amplitudes $x\equiv \alpha/\sqrt{n_0}$, $y\equiv \eta/(\kappa \sqrt{n_0})$, where $n_0=(\gamma \gamma_p)/(4g_0^2)$ is the saturation photon number. Eq.~(\ref{anm}) then becomes,
\begin{eqnarray}
y \hspace{-0.05in} &=& \hspace{-0.05in} x \Bigg(1 +\frac{N C_0}{4} \int d\delta P(\delta) \Bigg\{ \frac{1-i(\Delta+\delta)/\gamma_p}{1+\frac{(\Delta+\delta)^2}{\gamma_p^2}+\frac{|x|^2}{4} \left(1+ \xi^+\right) }
\nonumber
\\
&& +\frac{1- i(\Delta-\delta)/\gamma_p}{1+\frac{(\Delta-\delta)^2}{\gamma_p^2}+\frac{|x|^2}{4} \left(1+\xi^- \right) } \Bigg\} \Bigg),
\label{al0}
\end{eqnarray}
where $C_0 \equiv \frac{g_0^2}{\kappa \gamma_p}$, and $\xi^{\pm} =\frac{\gamma_p^2+(\Delta \pm \delta)^2}{\gamma_p^2+(\Delta \mp \delta)^2} $.

It is interesting to consider the relation corresponding to Eq.~(\ref{al0}) for a ring cavity system that has a field traveling in only one direction, assuming equal intracavity power. The corresponding expression is given by,

\begin{equation}
y = x \Bigg(1 + \frac{N C_0}{2} \int d\delta P(\delta) \frac{1-i(\Delta+\delta)/\gamma_p}{1+\frac{(\Delta+\delta)^2}{\gamma_p^2}+\frac{|x|^2}{2}} \Bigg).
\end{equation}

\begin{figure}
\begin{center}
	\includegraphics[scale=0.37] {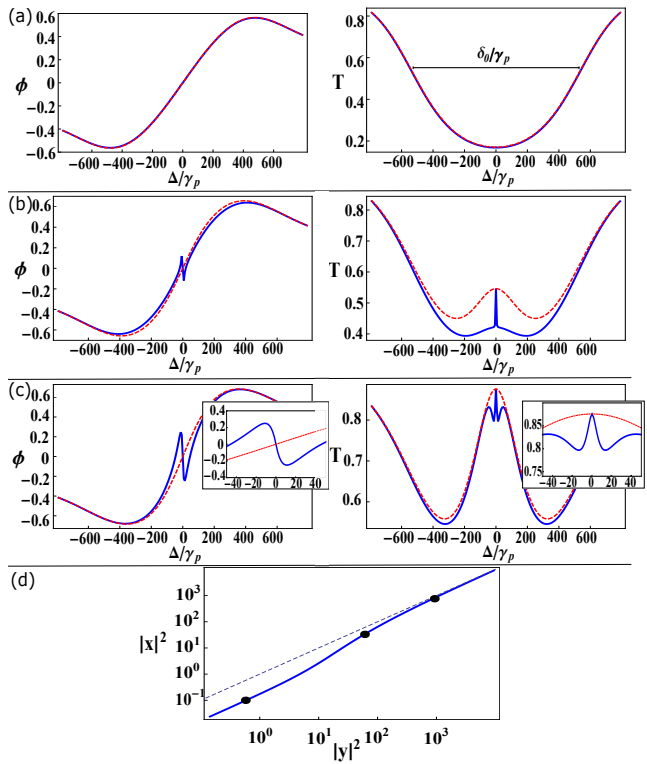}
\end{center}
\caption{(color online). Development of the extra absorption and dispersion features in the transmission (T) and phase shift ($\phi$), as the input intensity is increased. Blue (dark gray) solid curves are for a standing wave cavity, red (light gray) dashed are for a traveling wave cavity. For all plots, $N C_0 = 600$ and $\delta_0/\gamma_p=260$, which in the case of $^{88}$Sr, corresponds to a temperature of $\sim$ 15 $mK$. (a) $|y|^2=5\times10^{-1}$, (b) $|y|^2=6\times10^{1}$, (c) $|y|^2=9\times10^{2}$; The inset is zoomed in to emphasize the central features.  (d) Input vs intracavity intensity at resonance. The black dots label the input and intracavity intensities of (a) (b) and (c), and the dashed line is $|y|^2 = |x|^2$ for reference. Note that there is no bistability.}
\label{TransAndPhaseDifferentPLabeled}
\end{figure}

In the experiment described in \cite{JanPRL}, the measured observables are the cavity transmitted power $T\equiv |x/y|^2 $ and transmitted phase shift $\phi \equiv arg(x/y)$ of the intracavity light relative to the input light. Fig.~\ref{TransAndPhaseDifferentPLabeled} shows that the presence of $\xi^{\pm}$ in Eq.~(\ref{al0}) results in extra absorption and dispersive features (blue solid) around resonance in the transmission and phase shift, as compared to a ring cavity field (red dashed) where $\xi^{\pm}=0$.

These extra absorption and dispersive features are caused by the following: The distribution of atomic velocities results in a different Doppler frequency shift of the light for each atomic velocity class, so that each velocity class will be resonant at a different detuning $\Delta$. When $|\Delta| \ll \gamma_p |x|$, the resonant velocity class of the atoms is interrogated by both components of the standing wave field, whereas when $|\Delta| \gg \gamma_p |x|$, the resonant velocity class of atoms is interrogated by only one component of the standing wave field. Thus, there is an increased saturation in atomic absorption, with a corresponding saturated dispersive feature, for $|\Delta| \ll \gamma_p |x|$. These sharp features are absent from the traveling wave cavity situation where there is only one propagating field.

As seen in Fig.~\ref{TransAndPhaseDifferentPLabeled} (a), for low input intensities, there is no atomic saturation at any detuning. In Fig.~\ref{TransAndPhaseDifferentPLabeled} (b), the atoms in the velocity class around resonance are saturated by both components of the field, and the velocity classes away from resonance are saturated by only a single component of the standing wave field. Therefore, the features caused by the two component saturation and the features caused by the single component saturation are clearly able to be distinguished. It can be seen in Fig.~\ref{TransAndPhaseDifferentPLabeled} (c) that as the amount of saturation becomes large, the central feature becomes power broadened, but is still identifiable. Fig.~\ref{TransAndPhaseDifferentPLabeled} (d) shows the intracavity intensity for a given input intensity with the values used in Figs.~\ref{TransAndPhaseDifferentPLabeled} (a-c) labeled by black dots.

\section{Shot noise limited laser stabilization}
For laser stabilization, the central part of the phase response close to the atomic resonance serves as an error signal, see Fig.~\ref{TransAndPhaseDifferentPLabeled} (c), and allows for the generation of a feedback signal to the laser frequency. Through the photodetector, the error signal is converted to a measurable photocurrent. Any photocurrent measured by the photo detector should, in principle be zeroed by an ideal feedback loop of infinite bandwidth to the laser frequency. 

To determine the potential phase stability that could be achievable using our system, we consider the shot noise limited stabilization linewidth assuming a strong local oscillator, as derived in \cite{PhysRevA.84.063813},

\begin{equation}
\Delta \nu  = \frac{ \hbar \omega}{8 \pi \varepsilon P_{sig} \left( \frac{ \partial \phi}{\partial \Delta} \right)^2} = \frac{ C_0}{4 \pi \varepsilon\gamma |x|^2 \left( \frac{ \partial \phi}{\partial \Delta} \right)^2},
\label{linewidth}
\end{equation}
where $P_{sig}$ is the signal power, $\varepsilon$ is the photo-detector efficiency and $(\frac{ \partial \phi}{\partial \Delta})$ is the dimensionless phase slope at resonance $\Delta=0$. In the case that FM spectroscopy, such as NICEOHMS \cite{JanPRL}, is used for the detection of the absorption or cavity transmitted phase, Eq.~(\ref{linewidth}) must be modified as follows. In general, in a configuration where sidebands are applied at the free spectral range of the cavity, Eq.~(\ref{linewidth}) must be multiplied with $(1+P_{sig}/2P_{sideband})$, where $P_{sideband}$ is the sideband power. 

Eq.~(\ref{linewidth}) is the smallest, and hence the phase is most stable, when the product of the slope around resonance and the intracavity intensity is as large as possible. The optimal input intensity, which allows this product to be as large as possible, is the value used in Fig.~\ref{TransAndPhaseDifferentPLabeled} (c), and is labeled by the black dot at $|y|^2\approx10^3$ in Fig.~\ref{TransAndPhaseDifferentPLabeled} (d).

To achieve a quantitative agreement between theory and experiment \cite{JanPRL}, higher orders in $l$ in Eq.~(\ref{anm}) must be included. These higher order terms correspond to Doppleron resonances, i.e. multi-photon scattering processes between the atoms and cavity mode.

To study the importance of these higher order Dopplerons, we calculate the linewidth from Eq.~(\ref{linewidth}) at the optimum input power while varying the order of $l$ at which the truncation occurs.

\begin{figure}[t]
\begin{center}
	\includegraphics[scale =0.5] {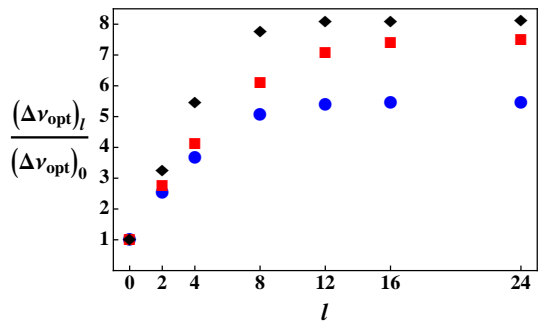}
\end{center}
\caption{(color online). Linewidth at the optimum $|y|^2$, which gives the smallest linewidth, as a function of included orders of $l$, normalized by the $l=0$ linewidth.
Blue circles: $NC_0=600$, $\delta_0/\gamma_p=260$, $|y|^2=1270$.
Red squares: $NC_0=6000$, $\delta_0/\gamma_p=260$, $|y|^2=7700$.
Black diamonds: $NC_0=600$, $\delta_0/\gamma_p=80$, $|y|^2=850$.}
\label{DoppleronComparison}
\end{figure}

Fig.~\ref{DoppleronComparison} shows the dependence of the linewidth at the optimum input intensity on the order of $l$ at which the truncation occurs for 3 different sets of parameters. The first set, shown in blue circles, converges by $l=12$. The linewidth calculated with up to $l=12$ included before truncation is around 5 times larger than the linewidth with only $l=0$ included. This shows that Doppleron effects are crucial to include for a correct quantitative analysis of this system.

Shown in red squares, we increase the number of atoms by a factor of 10, and again calculate the linewidth as a function of the order of $l$ at which the truncation occurs. As a result of N being increased, the optimum value of $|y|^2$ is also increased. Convergence occurs around $l=16$. Now, the difference between the converged linewidth and the linewidth with only $l=0$ included has increased by a factor of around 2. This demonstrates that as $N$ is increased, higher order Dopplerons play an increasing role.

We also decrease the temperature by a factor of 10, and again calculate the linewidth at the new optimum value of $|y|^2$, as shown by the black diamonds. Even though the optimum $|y|^2$ occurs at a lower value, there is still an increase in the difference between the converged linewidth and the $l=0$ linewidth. Convergence occurs around $l=12$. This shows that as the temperature is decreased, higher order Dopplerons also play an increasing role.

\begin{figure}[t]
\begin{center}
	\hspace{5mm} \includegraphics[scale =0.40]{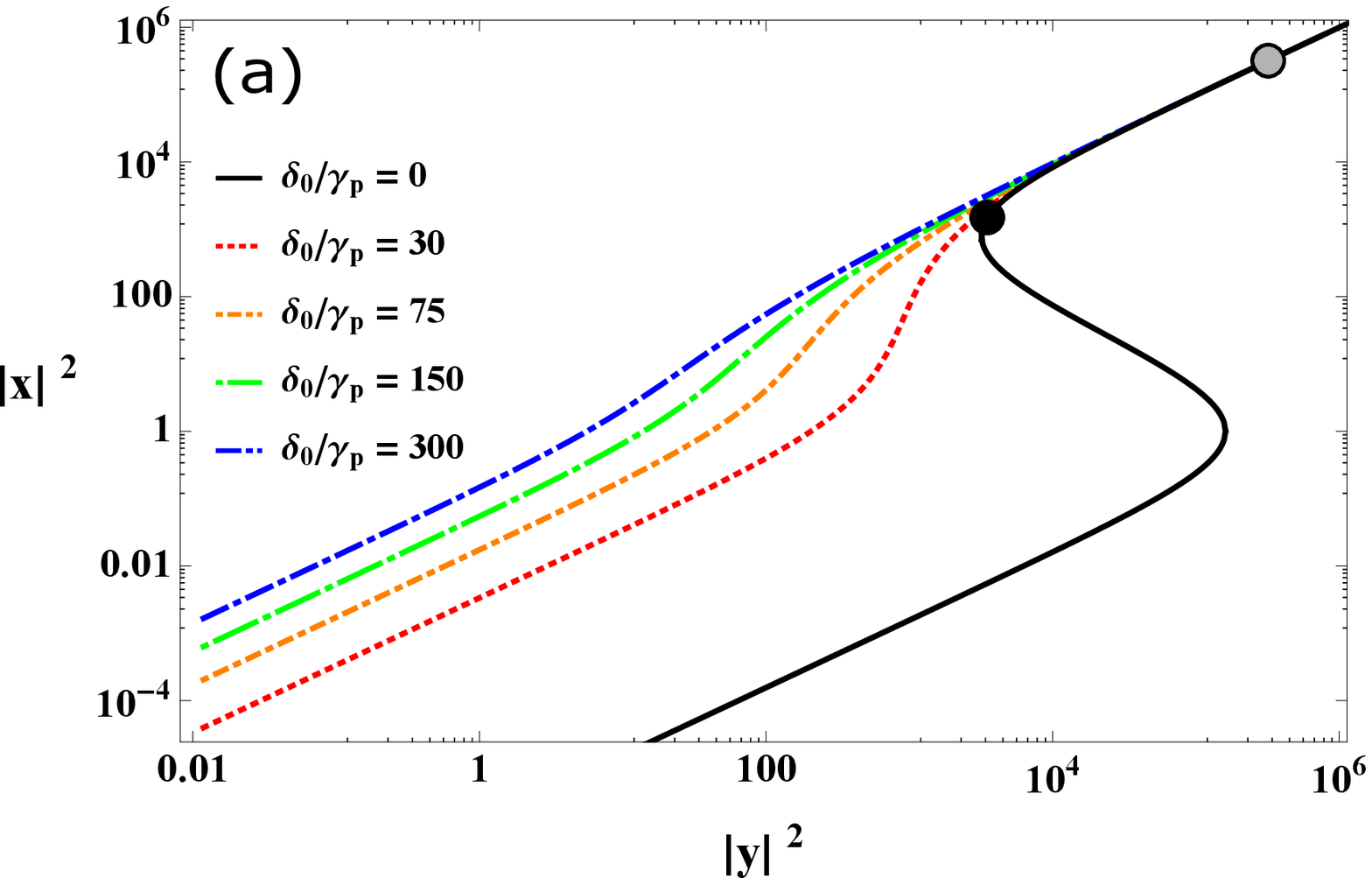}\\
	\hspace{4mm} \includegraphics[scale =0.55]{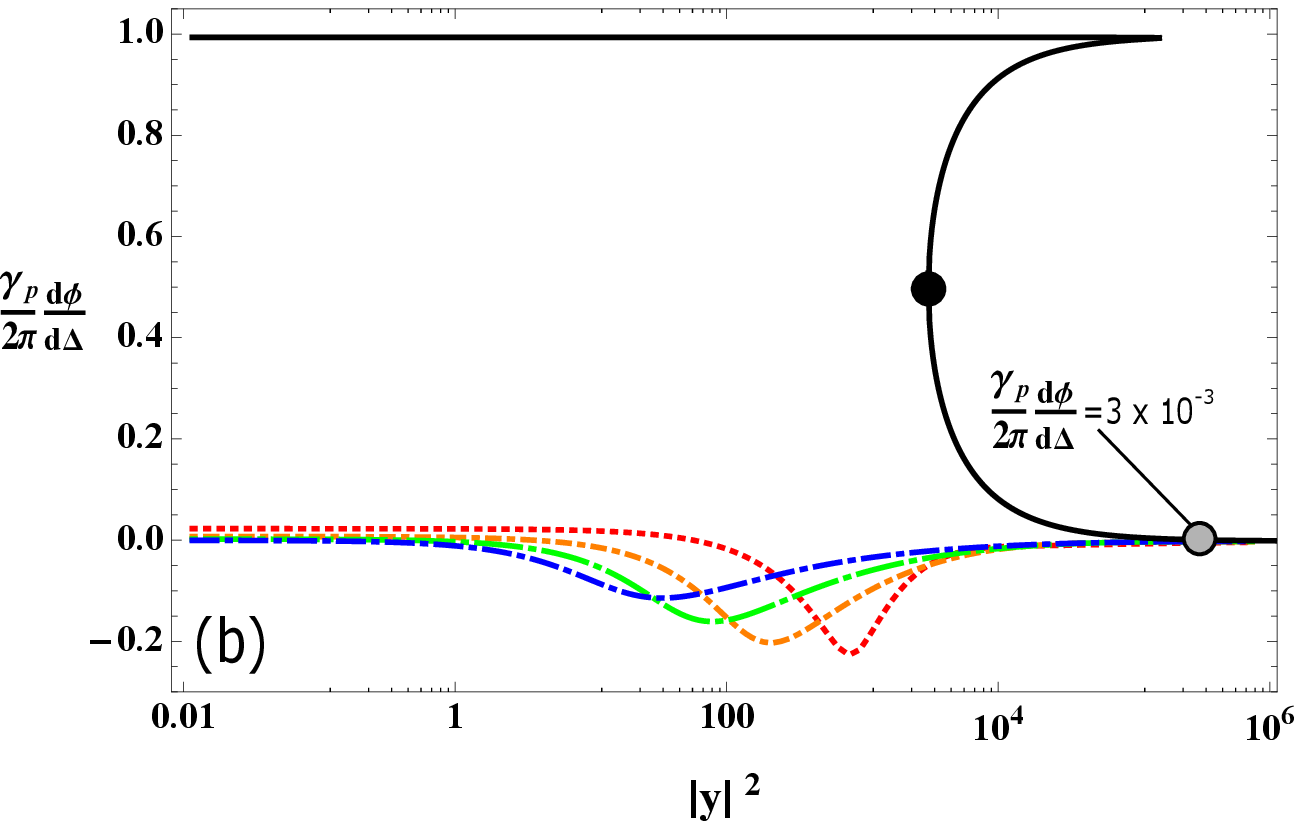}\\
	\includegraphics[scale =0.44]{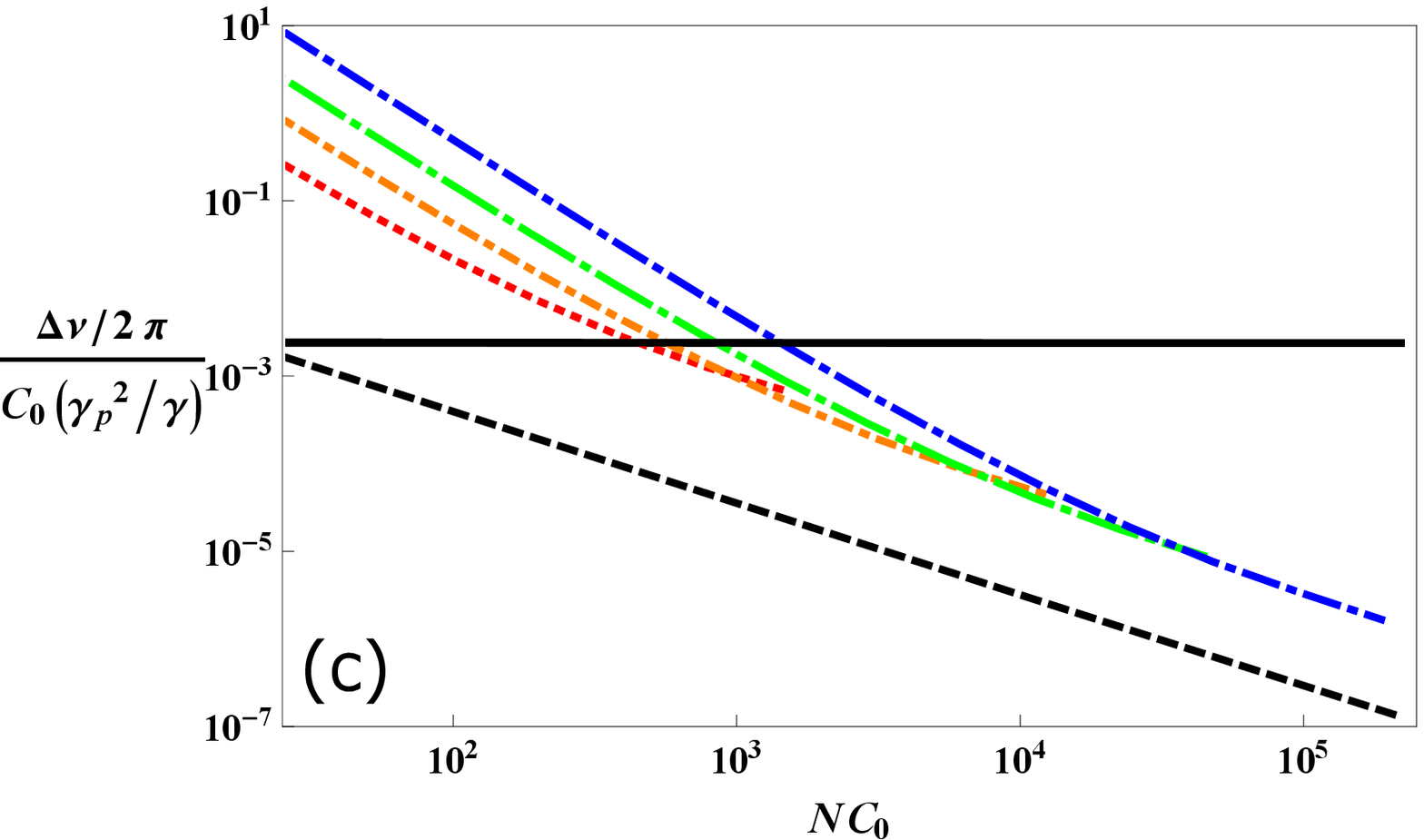}
\end{center}
\caption{(color online). (a) Scaled input ($|x|^2$) vs intracavity ($|y|^2$) intensities for several temperatures with $N C_0=800$ (b) Slope at resonance for several temperatures with $N C_0=800$ (c) Stabilization linewidth as a function of $N C_0$ for several temperatures. For $\delta_0 / \gamma_p = 0$, black dashed is calculated at the fixed input intensity $|y|^2=4NC_0$; black solid is calculated at the fixed input intensity $|y|^2=(NC_0)^2/4$. For $\delta_0 / \gamma_p \neq 0$, the linewidth is calculated with the input intensity fixed to the optimum value that gives the smallest linewidth. In the case of $^{88}$Sr, $\delta_0 / \gamma_p = 150$ corresponds to a temperature of $\sim$ 5 $mK$.}
\label{PinPoutAndSlopeAll}
\end{figure}

We next study the effect of optical bistability, and its effects on the optimum input intensity. As can be seen in Fig.~\ref{PinPoutAndSlopeAll} (a), when $\delta_0 / \gamma_p=0$ there is a bistability in input ($|y|^2$) vs intracavity ($|x|^2$) intensities. As the temperature is increased, the bistability becomes less pronounced, until eventually it disappears entirely, which can be seen in the  $\delta_0 / \gamma_p=30$ curve.

The slope at resonance, which can be seen in Fig.~\ref{PinPoutAndSlopeAll} (b), has very a different behavior for the zero and non zero temperature cases. For $\delta_0 / \gamma_p=0$, the slope is largest on the lower branch, decreases slightly in the bistable region, and then drops exponentially to zero as the intensity is increased on the upper branch. When the temperature is increased, the bistability disappears, and a dispersion feature with negative slope appears. Then the slope goes to a maximally negative value before power broadening eventually causes it to increase to zero.

\begin{table*}[t]
\begin{ruledtabular}
\begin{tabular}{p{3 cm}|c|c|c|c|c|c||c|c}
Transision & $\lambda$ & $\gamma/2\pi$ & $F$ & $N$ & $N C_0$ & Temperature & $P_{\textrm{\footnotesize{in,opt}}}$ & $\Delta \nu$\\
\hline\hline
\multirow{2}*{$^{171}$Yb $^{1}$S$_{0}\rightarrow^{3}$P$_{1}$} & \multirow{2}*{$556\,\textrm{nm}$} & \multirow{2}*{$182\,\textrm{kHz}$} & $250$ & $2.5\cdot 10^{7}$ & $374$ & \multirow{2}*{$6.5\, \textrm{mK}$} & $128\,\mu\textrm{W}$ & $3.3\,\textrm{Hz}$\\
& & & $1000$ & $5.0\cdot 10^{7}$ & $2991$ & & $312\,\mu\textrm{W}$ & $207\,\textrm{mHz}$\\
\hline
\multirow{2}*{$^{40}$Ca $^{1}$S$_{0}\rightarrow^{3}$P$_{1}$} & \multirow{2}*{$657\,\textrm{nm}$} & \multirow{2}*{$400\,\textrm{Hz}$} & $250$ & $2.5\cdot 10^{7}$ & $522$ & \multirow{2}*{$1.7\,\textrm{mK}$} & $5.5\,\textrm{nW}$ & $322\,\textrm{mHz}$\\
& & & $1000$ & $5.0\cdot 10^{7}$ & $4176$ & & $2.2\,\textrm{nW}$ & $20\,\textrm{mHz}$\\
\hline
\multirow{2}*{$^{24}$Mg $^{1}$S$_{0}\rightarrow^{3}$P$_{1}$} & \multirow{2}*{$457\,\textrm{nm}$} & \multirow{2}*{$34\,\textrm{Hz}$} & $250$ & $2.5\cdot 10^{7}$ & $253$ & \multirow{2}*{$3.0\,\textrm{mK}$} & $0.5\,\textrm{nW}$ &$8.1\,\textrm{Hz}$\\
& & & $1000$ & $5.0\cdot 10^{7}$ & $2021$ &  & $0.2\,\textrm{nW}$ & $500\,\textrm{mHz}$\\
\hline
\multirow{2}*{$^{88}$Sr $^{1}$S$_{0}\rightarrow^{3}$P$_{1}$} & \multirow{2}*{$689\,\textrm{nm}$} & \multirow{2}*{$7.6\,\textrm{kHz}$} & $250$ & $2.5\cdot 10^{7}$ & $574$ & \multirow{2}*{$3.0\,\textrm{mK}$} & $47\,\textrm{nW}$ & $102\,\textrm{mHz}$\\
& & & $1000$ & $5.0\cdot 10^{7}$ & $4593$ & & $84\,\textrm{nW}$ & $6.8\,\textrm{mHz}$\\
\end{tabular}
\end{ruledtabular}
\caption{Shot noise limited linewidths $\Delta \nu$ estimated theoretically for a number of $^{1}$S$_{0}\rightarrow^{3}$P$_{1}$ intercombination lines at experimentally realizable parameters: transition wavelength $\lambda$, natural linewidth $\gamma$, atomic sample temperature as reported in \cite{MgPaper,YbPaper,CaPaper} and optimal cavity input power $P_{\textrm{\tiny{in,opt}}}$. The critical temperature at which bistability disappears for $N C_0=4593$, the largest value of $N C_0$ here considered, is $60\, \mu K$, which is well below any temperatures considered. The ratio of the carrier and sideband power is chosen to  $\frac{P_{\textrm{\tiny{sig}}}}{2P_{\textrm{\tiny{sideband}}}}=1$ in all cases. Throughout the calculation we have assumed a prestabilized probe laser decoherence of $\gamma_{\textrm{\tiny{laser}}}/2\pi = 2.0\,\textrm{kHz}$. The shot noise limited linewidths are calculated for systems with empty cavity finesse of $F=250$ and $N=2.5\cdot 10^{7}$ atoms overlapping the cavity mode, and an improved system with $F=1000$ and $N=5.0\cdot 10^{7}$.}
\label{linewidthtable}
\end{table*}

The disappearance of the bistability is important for achieving the highest possible degree of stabilization. In the zero temperature case, the optimal combination of $\left| x \right| ^2$ and $\frac{ \partial \phi}{\partial \Delta}$ to give the smallest $\Delta \nu$ occurs on the far left side of the upper branch, when the input intensity has the fixed value $|y|^2=4NC_0$, labeled with the black dot in Fig.~\ref{PinPoutAndSlopeAll} (a) and (b). However, this value of intensity is in the bistable region, so the system is dynamically unstable when the full quantum dynamics are accounted for. If the tunneling rate between the different branches is not small, we are constrained to work at input intensities that are above the bistable region where the achievable frequency precision is much less.  The grey dot in Fig.~\ref{PinPoutAndSlopeAll} (a) and (b) is at the far right side of the bistable region, at the fixed input intensity value $|y|^2=(NC_0)^2/4$. The stabilization linewidth is orders of magnitude worse here, because the slope is so small. Nonetheless, working above the bistability allows for shot noise limited stabilization line widths of $\sim1$ mHz \cite{PhysRevA.84.063813}.

Fig.~\ref{PinPoutAndSlopeAll} (c) shows the stabilization linewidth as a function of $NC_0$ for several temperatures. In the $\delta_0 / \gamma_p = 0$ limit of zero temperature, the black dashed curve corresponds to the input intensity corresponding to the black dot, and the black solid curve corresponds to the input intensity of the gray dot. When the temperature is increased sufficiently, the bistability disappears. Then, there are no longer any regions of dynamic instability and therefore no restrictions on input intensity. Each of the $\delta_0 / \gamma_p \neq 0$ cases in this plot are calculated at their respective optimal input intensity, and stop at a critical value of $NC_0$, where the bistability reappears. It follows as reciprocal information that for a given value of $NC_0$, there exists a critical temperature at which the bistability disappears. Therefore, by using Fig.~\ref{PinPoutAndSlopeAll} (c)  to note the values of temperature and $NC_0$ at which the bistability disappears for the different curves, it is seen that the critical temperature increases as $NC_0$ increases.

In general, a lower temperature will yield a smaller linewidth. However, because one must avoid working in a region of optical bistability, the optimal shot noise limited linewidth for certain values of  $NC_0$ and $\delta_0 / \gamma_p$ in which there is no bistability can actually be smaller than the $\delta_0 / \gamma_p =0$ linewidth which is constrained above the bistability.

\section{Atomic systems}

In order to evaluate the performance of the cavity QED stabilization method presented in this work, we have estimated the shot noise limited linewidth using equation~(\ref{linewidth}) for for a number of different atomic systems. For each element we assume the atomic sample laser to be cooled close to the  Doppler limit with reference to temperatures obtained experimentally in \cite{YbPaper,MgPaper,CaPaper,JanPRL}, see Table~\ref{linewidthtable}. These temperatures are all above the critical temperature where bi-stability disappears by at least an order of magnitude. 

A realistic atom number overlapping the cavity mode has been estimated to be $N=2.5\cdot 10^{7}$ and has been used for all the elements. For several elements, larger atom numbers have been reported in the literature which makes our estimates somewhat conservative \cite{YbPaper,MgPaper,CaPaper,JanPRL}. Cavity dimensions are based on the experimental values obtained in \cite{JanPRL}. The cavity waist diameter is chosen to be $1.0\,\textrm{mm}$ and the empty cavity finesse to be $F=250$ and $F=1000$, corresponding to an empty cavity decay rate of $\kappa/2\pi = 2.0\cdot 10^{6}\,\textrm{Hz}$ and $\kappa/2\pi = 0.5\cdot 10^{6}\,\textrm{Hz}$ respectively. In each case, the laser decoherence has been assumed to be $\gamma_{\textrm{\small{laser}}}=2\pi\cdot 2.0\,\textrm{kHz}$.  For each element in Table~\ref{linewidthtable}, we have used the laser power that corresponds to the minimum shot noise limited linewidth.

Whereas stabilizing a probe laser on the $^{1}$S$_{0}\rightarrow^{3}$P$_{1}$ transition of $^{24}$Mg may be slightly more experimentally challenging compared to the other elements in Table~\ref{linewidthtable}, the stabilization of a probe laser on the $^{1}$S$_{0}\rightarrow^{3}$P$_{1}$ transition of $^{88}$Sr with cavity finesse $F=250$ and $N=2.5\cdot 10^{7}$ promises shot noise limited linewidth of $107\,\textrm{mHz}$. This linewidth is already comparable with the linewidths of the current state-of-the-art frequency stabilized lasers \cite{JanPRL, Martin}.

The shot noise limited linewidth may be reduced further by increasing the value of $NC_{0}$. This can be achieved by increasing the atom number or the cavity finesse. The shot noise limited linewidth, $\Delta\nu$, is inversely proportional to the square of the atom number according to equation~(\ref{linewidth}), as the phase slope at resonance, $\frac{ \partial \phi}{\partial \Delta}$, depends linearly on the atom number $N$. A similar scaling can be achieved by increasing the finesse while keeping the classical drive amplitude $\eta$ constant.
For $^{88}$Sr, the laser linewidth may be reduced to $\Delta\nu=6.8\,\textrm{mHz}$ by only increasing the cavity finesse to $F=1000$ and increasing the atom number by a factor of two to $N=5.0\cdot 10^{7}$, see Table~\ref{linewidthtable}.

A cavity imposes a delay on the error signal possibly degrading the feedback to the laser. Generally, the optical cavity effectively acts as a low pass filter with a cut-off frequency given by $f_{cut} = c/(4LF)$ \cite{2002PhLA..305..239R}, where $c$ is the speed of light, $L$ is the cavity length and $F$ the finesse of the cavity. In the case studied in this paper we find the cavity belonging to the so-called bad cavity limit, where the cavity linewidth $\Gamma_c$ is significantly larger than the the atomic linewidth $\gamma_a$. For a typical cavity of length 10 cm, operated in this domain, the finesse ranges from 100-1000 and corresponds to cut-off frequencies in the MHz range. For practical purposes this will not pose any significant delay in the servo loop.           

The experimental proof-of-principle demonstrated in \cite{JanPRL} is operated in a cyclic fashion, on a time scale similar to that of state-of-the-art optical lattice clocks. In these clocks, the preparation of cold atoms takes of order 0.5 - 1 seconds, before they are interrogated by the clock laser. This leaves the probe (clock) laser uncorrected during the dead time period.  All such systems suffer from the so-called optical Dick effect \cite{Dick,westergaard2010minimizing} where the frequency noise of the clock laser is aliased into the sampled signal and ends up contaminating the clock stability. Possible ways to decrease the Dick effect is to increase the probe period by transferring the atoms to an optical lattice or by reducing the initial clock laser instability. With a laser noise level similar to that achieved by optical lattice clocks, and a duty cycle of about one second, we would be limited to stability at the $10^{-15}$ level \cite{5977827}. Compared to the estimated shot noise limited line widths presented in Table~\ref{linewidthtable}, this corresponds to the low finesse cases. Pushing the limit, as shown by the slightly higher finesse of 1000, would require the atoms to be loaded into an optical lattice, or require a reduction of the duty cycle by a factor 10 to 100 ms in the experiment presented in \cite{JanPRL}. This may be achieved with an optimized experimental loading rate as shown in \cite{AtomFlux}.

Alternatively, one may use a cold bright atomic beam as a source of atoms to be interrogated in the cavity, rather than atoms prepared in a MOT and then released. Using experimentally achieved numbers from \cite{AtomFlux}, we have estimated signals comparable to that achieved in \cite{JanPRL}, but now allowing for a continuous interrogation. This seems to be a very tantalizing and promising approach for these systems. 
 
\section{Conclusion}
We have seen that thermal atoms in a standing wave cavity field exhibit additional phenomena that were not observed when considering a frozen arrangement of atoms. Specifically, when the detuning of the laser and atoms is less than the power broadened linewidth, the system interacts with both components of the standing wave field. This causes new absorptive and dispersive features in the observables, which are the features that can be used as an error signal for frequency stabilization. Multi-photon scattering processes due to Doppleron resonances cause the stabilization linewidth to increase. This effect becomes more dominant as the collective cooperativity $NC_0$ is increased, and as the temperature is decreased. A system with sufficient $NC_0$ and no atomic motion will exhibit optical bistability. Atomic motion may cause this bistability to disappear. When optical bistability occurs, one is generally restricted to using an input intensity that lies outside of the bistable region in order to prevent hopping between semiclassical solutions. When there is no optical bistability, there are no dynamically unstable regions, so that no such restrictions on input intensity are necessary, allowing the optimally smallest stabilization linewidth to be achieved.

The authors acknowledge valuable conversations with G. S. Agarwal. This work has been supported by JILA-NSF-PFC-1125844, DARPA, QuASAR, NIST and The Danish research council and ESA contract No. 4000108303/13/NL/PA-NPI272-2012.

\bibliographystyle{apsrev4-1}
\bibliography{DoppleronPaper}

\begin{thebibliography}{34}%
\makeatletter
\providecommand \@ifxundefined [1]{%
 \@ifx{#1\undefined}
}%
\providecommand \@ifnum [1]{%
 \ifnum #1\expandafter \@firstoftwo
 \else \expandafter \@secondoftwo
 \fi
}%
\providecommand \@ifx [1]{%
 \ifx #1\expandafter \@firstoftwo
 \else \expandafter \@secondoftwo
 \fi
}%
\providecommand \natexlab [1]{#1}%
\providecommand \enquote  [1]{``#1''}%
\providecommand \bibnamefont  [1]{#1}%
\providecommand \bibfnamefont [1]{#1}%
\providecommand \citenamefont [1]{#1}%
\providecommand \href@noop [0]{\@secondoftwo}%
\providecommand \href [0]{\begingroup \@sanitize@url \@href}%
\providecommand \@href[1]{\@@startlink{#1}\@@href}%
\providecommand \@@href[1]{\endgroup#1\@@endlink}%
\providecommand \@sanitize@url [0]{\catcode `\\12\catcode `\$12\catcode
  `\&12\catcode `\#12\catcode `\^12\catcode `\_12\catcode `\%12\relax}%
\providecommand \@@startlink[1]{}%
\providecommand \@@endlink[0]{}%
\providecommand \url  [0]{\begingroup\@sanitize@url \@url }%
\providecommand \@url [1]{\endgroup\@href {#1}{\urlprefix }}%
\providecommand \urlprefix  [0]{URL }%
\providecommand \Eprint [0]{\href }%
\@ifxundefined \urlstyle {%
  \providecommand \doi  [0]{\begingroup \@sanitize@url \@doi}%
  \providecommand \@doi [1]{\endgroup \@@startlink {\doibase
  #1}doi:\discretionary {}{}{}#1\@@endlink }%
}{%
  \providecommand \doi  [0]{doi:\discretionary{}{}{}\begingroup
  \urlstyle{rm}\Url }%
}%
\providecommand \doibase [0]{http://dx.doi.org/}%
\providecommand \Doi [0]{\begingroup \@sanitize@url \@Doi }%
\providecommand \@Doi  [1]{\endgroup\@@startlink{\doibase#1}\@@Doi}%
\providecommand \@@Doi [1]{#1\@@endlink}%
\providecommand \selectlanguage [0]{\@gobble}%
\providecommand \bibinfo  [0]{\@secondoftwo}%
\providecommand \bibfield  [0]{\@secondoftwo}%
\providecommand \translation [1]{[#1]}%
\providecommand \BibitemOpen [0]{}%
\providecommand \bibitemStop [0]{}%
\providecommand \bibitemNoStop [0]{.\EOS\space}%
\providecommand \EOS [0]{\spacefactor3000\relax}%
\providecommand \BibitemShut  [1]{\csname bibitem#1\endcsname}%
\bibitem [{\citenamefont {Le~Targat}\ \emph {et~al.}(2013)\citenamefont
  {Le~Targat}, \citenamefont {Lorini}, \citenamefont {Le~Coq}, \citenamefont
  {Zawada}, \citenamefont {Guéna}, \citenamefont {Abgrall}, \citenamefont
  {Gurov}, \citenamefont {Rosenbusch}, \citenamefont {Rovera}, \citenamefont
  {Nagórny}, \citenamefont {Gartman}, \citenamefont {Westergaard},
  \citenamefont {Tobar}, \citenamefont {Lours}, \citenamefont {Santarelli},
  \citenamefont {Clairon}, \citenamefont {Bize}, \citenamefont {Laurent},
  \citenamefont {Lemonde},\ and\ \citenamefont {Lodewyck}}]{LeTargat}%
  \BibitemOpen
  \bibfield  {author} {\bibinfo {author} {\bibfnamefont {R.}~\bibnamefont
  {Le~Targat}}, \bibinfo {author} {\bibfnamefont {L.}~\bibnamefont {Lorini}},
  \bibinfo {author} {\bibfnamefont {Y.}~\bibnamefont {Le~Coq}}, \bibinfo
  {author} {\bibfnamefont {M.}~\bibnamefont {Zawada}}, \bibinfo {author}
  {\bibfnamefont {J.}~\bibnamefont {Guéna}}, \bibinfo {author} {\bibfnamefont
  {M.}~\bibnamefont {Abgrall}}, \bibinfo {author} {\bibfnamefont
  {M.}~\bibnamefont {Gurov}}, \bibinfo {author} {\bibfnamefont
  {P.}~\bibnamefont {Rosenbusch}}, \bibinfo {author} {\bibfnamefont {D.~G.}\
  \bibnamefont {Rovera}}, \bibinfo {author} {\bibfnamefont {B.}~\bibnamefont
  {Nagórny}}, \bibinfo {author} {\bibfnamefont {R.}~\bibnamefont {Gartman}},
  \bibinfo {author} {\bibfnamefont {P.~G.}\ \bibnamefont {Westergaard}},
  \bibinfo {author} {\bibfnamefont {M.~E.}\ \bibnamefont {Tobar}}, \bibinfo
  {author} {\bibfnamefont {M.}~\bibnamefont {Lours}}, \bibinfo {author}
  {\bibfnamefont {G.}~\bibnamefont {Santarelli}}, \bibinfo {author}
  {\bibfnamefont {A.}~\bibnamefont {Clairon}}, \bibinfo {author} {\bibfnamefont
  {S.}~\bibnamefont {Bize}}, \bibinfo {author} {\bibfnamefont {P.}~\bibnamefont
  {Laurent}}, \bibinfo {author} {\bibfnamefont {P.}~\bibnamefont {Lemonde}}, \
  and\ \bibinfo {author} {\bibfnamefont {J.}~\bibnamefont {Lodewyck}},\ }\href
  {http://dx.doi.org/10.1038/ncomms3109} {\bibfield  {journal} {\bibinfo
  {journal} {Nat. Commun.},\ }\textbf {\bibinfo {volume} {4}} (\bibinfo {year}
  {2013})}\BibitemShut {NoStop}%
\bibitem [{\citenamefont {Huntemann}\ \emph {et~al.}(2014)\citenamefont
  {Huntemann}, \citenamefont {Lipphardt}, \citenamefont {Tamm}, \citenamefont
  {Gerginov}, \citenamefont {Weyers},\ and\ \citenamefont {Peik}}]{Huntemann}%
  \BibitemOpen
  \bibfield  {author} {\bibinfo {author} {\bibfnamefont {N.}~\bibnamefont
  {Huntemann}}, \bibinfo {author} {\bibfnamefont {B.}~\bibnamefont
  {Lipphardt}}, \bibinfo {author} {\bibfnamefont {C.}~\bibnamefont {Tamm}},
  \bibinfo {author} {\bibfnamefont {V.}~\bibnamefont {Gerginov}}, \bibinfo
  {author} {\bibfnamefont {S.}~\bibnamefont {Weyers}}, \ and\ \bibinfo {author}
  {\bibfnamefont {E.}~\bibnamefont {Peik}},\ }\Doi
  {10.1103/PhysRevLett.113.210802} {\bibfield  {journal} {\bibinfo  {journal}
  {Phys. Rev. Lett.},\ }\textbf {\bibinfo {volume} {113}},\ \bibinfo {pages}
  {210802} (\bibinfo {year} {2014})}\BibitemShut {NoStop}%
\bibitem [{\citenamefont {Godun}\ \emph {et~al.}(2014)\citenamefont {Godun},
  \citenamefont {Nisbet-Jones}, \citenamefont {Jones}, \citenamefont {King},
  \citenamefont {Johnson}, \citenamefont {Margolis}, \citenamefont {Szymaniec},
  \citenamefont {Lea}, \citenamefont {Bongs},\ and\ \citenamefont
  {Gill}}]{Godun}%
  \BibitemOpen
  \bibfield  {author} {\bibinfo {author} {\bibfnamefont {R.~M.}\ \bibnamefont
  {Godun}}, \bibinfo {author} {\bibfnamefont {P.~B.~R.}\ \bibnamefont
  {Nisbet-Jones}}, \bibinfo {author} {\bibfnamefont {J.~M.}\ \bibnamefont
  {Jones}}, \bibinfo {author} {\bibfnamefont {S.~A.}\ \bibnamefont {King}},
  \bibinfo {author} {\bibfnamefont {L.~A.~M.}\ \bibnamefont {Johnson}},
  \bibinfo {author} {\bibfnamefont {H.~S.}\ \bibnamefont {Margolis}}, \bibinfo
  {author} {\bibfnamefont {K.}~\bibnamefont {Szymaniec}}, \bibinfo {author}
  {\bibfnamefont {S.~N.}\ \bibnamefont {Lea}}, \bibinfo {author} {\bibfnamefont
  {K.}~\bibnamefont {Bongs}}, \ and\ \bibinfo {author} {\bibfnamefont
  {P.}~\bibnamefont {Gill}},\ }\Doi {10.1103/PhysRevLett.113.210801} {\bibfield
   {journal} {\bibinfo  {journal} {Phys. Rev. Lett.},\ }\textbf {\bibinfo
  {volume} {113}},\ \bibinfo {pages} {210801} (\bibinfo {year}
  {2014})}\BibitemShut {NoStop}%
\bibitem [{\citenamefont {Schiller}\ \emph {et~al.}(2009)\citenamefont
  {Schiller}, \citenamefont {Tino}, \citenamefont {Gill}, \citenamefont
  {Salomon}, \citenamefont {Sterr}, \citenamefont {Peik}, \citenamefont
  {Nevsky}, \citenamefont {Görlitz}, \citenamefont {Svehla}, \citenamefont
  {Ferrari}, \citenamefont {Poli}, \citenamefont {Lusanna}, \citenamefont
  {Klein}, \citenamefont {Margolis}, \citenamefont {Lemonde}, \citenamefont
  {Laurent}, \citenamefont {Santarelli}, \citenamefont {Clairon}, \citenamefont
  {Ertmer}, \citenamefont {Rasel}, \citenamefont {Müller}, \citenamefont
  {Iorio}, \citenamefont {Lämmerzahl}, \citenamefont {Dittus}, \citenamefont
  {Gill}, \citenamefont {Rothacher}, \citenamefont {Flechner}, \citenamefont
  {Schreiber}, \citenamefont {Flambaum}, \citenamefont {Ni}, \citenamefont
  {Liu}, \citenamefont {Chen}, \citenamefont {Chen}, \citenamefont {Gao},
  \citenamefont {Cacciapuoti}, \citenamefont {Holzwarth}, \citenamefont
  {Heß},\ and\ \citenamefont {Schäfer}}]{Schiller}%
  \BibitemOpen
  \bibfield  {author} {\bibinfo {author} {\bibfnamefont {S.}~\bibnamefont
  {Schiller}}, \bibinfo {author} {\bibfnamefont {G.}~\bibnamefont {Tino}},
  \bibinfo {author} {\bibfnamefont {P.}~\bibnamefont {Gill}}, \bibinfo {author}
  {\bibfnamefont {C.}~\bibnamefont {Salomon}}, \bibinfo {author} {\bibfnamefont
  {U.}~\bibnamefont {Sterr}}, \bibinfo {author} {\bibfnamefont
  {E.}~\bibnamefont {Peik}}, \bibinfo {author} {\bibfnamefont {A.}~\bibnamefont
  {Nevsky}}, \bibinfo {author} {\bibfnamefont {A.}~\bibnamefont {Görlitz}},
  \bibinfo {author} {\bibfnamefont {D.}~\bibnamefont {Svehla}}, \bibinfo
  {author} {\bibfnamefont {G.}~\bibnamefont {Ferrari}}, \bibinfo {author}
  {\bibfnamefont {N.}~\bibnamefont {Poli}}, \bibinfo {author} {\bibfnamefont
  {L.}~\bibnamefont {Lusanna}}, \bibinfo {author} {\bibfnamefont
  {H.}~\bibnamefont {Klein}}, \bibinfo {author} {\bibfnamefont
  {H.}~\bibnamefont {Margolis}}, \bibinfo {author} {\bibfnamefont
  {P.}~\bibnamefont {Lemonde}}, \bibinfo {author} {\bibfnamefont
  {P.}~\bibnamefont {Laurent}}, \bibinfo {author} {\bibfnamefont
  {G.}~\bibnamefont {Santarelli}}, \bibinfo {author} {\bibfnamefont
  {A.}~\bibnamefont {Clairon}}, \bibinfo {author} {\bibfnamefont
  {W.}~\bibnamefont {Ertmer}}, \bibinfo {author} {\bibfnamefont
  {E.}~\bibnamefont {Rasel}}, \bibinfo {author} {\bibfnamefont
  {J.}~\bibnamefont {Müller}}, \bibinfo {author} {\bibfnamefont
  {L.}~\bibnamefont {Iorio}}, \bibinfo {author} {\bibfnamefont
  {C.}~\bibnamefont {Lämmerzahl}}, \bibinfo {author} {\bibfnamefont
  {H.}~\bibnamefont {Dittus}}, \bibinfo {author} {\bibfnamefont
  {E.}~\bibnamefont {Gill}}, \bibinfo {author} {\bibfnamefont {M.}~\bibnamefont
  {Rothacher}}, \bibinfo {author} {\bibfnamefont {F.}~\bibnamefont {Flechner}},
  \bibinfo {author} {\bibfnamefont {U.}~\bibnamefont {Schreiber}}, \bibinfo
  {author} {\bibfnamefont {V.}~\bibnamefont {Flambaum}}, \bibinfo {author}
  {\bibfnamefont {W.-T.}\ \bibnamefont {Ni}}, \bibinfo {author} {\bibfnamefont
  {L.}~\bibnamefont {Liu}}, \bibinfo {author} {\bibfnamefont {X.}~\bibnamefont
  {Chen}}, \bibinfo {author} {\bibfnamefont {J.}~\bibnamefont {Chen}}, \bibinfo
  {author} {\bibfnamefont {K.}~\bibnamefont {Gao}}, \bibinfo {author}
  {\bibfnamefont {L.}~\bibnamefont {Cacciapuoti}}, \bibinfo {author}
  {\bibfnamefont {R.}~\bibnamefont {Holzwarth}}, \bibinfo {author}
  {\bibfnamefont {M.}~\bibnamefont {Heß}}, \ and\ \bibinfo {author}
  {\bibfnamefont {W.}~\bibnamefont {Schäfer}},\ }\Doi
  {10.1007/s10686-008-9126-5} {\emph {\bibinfo {title} {Einstein Gravity
  Explorer–a medium-class fundamental physics mission}}},\ Vol.~\bibinfo
  {volume} {23}\ (\bibinfo  {publisher} {Springer Netherlands},\ \bibinfo
  {year} {2009})\ pp.\ \bibinfo {pages} {573--610}\BibitemShut {NoStop}%
\bibitem [{\citenamefont {Gu\'ena}\ \emph {et~al.}(2012)\citenamefont
  {Gu\'ena}, \citenamefont {Abgrall}, \citenamefont {Rovera}, \citenamefont
  {Rosenbusch}, \citenamefont {Tobar}, \citenamefont {Laurent}, \citenamefont
  {Clairon},\ and\ \citenamefont {Bize}}]{Guena}%
  \BibitemOpen
  \bibfield  {author} {\bibinfo {author} {\bibfnamefont {J.}~\bibnamefont
  {Gu\'ena}}, \bibinfo {author} {\bibfnamefont {M.}~\bibnamefont {Abgrall}},
  \bibinfo {author} {\bibfnamefont {D.}~\bibnamefont {Rovera}}, \bibinfo
  {author} {\bibfnamefont {P.}~\bibnamefont {Rosenbusch}}, \bibinfo {author}
  {\bibfnamefont {M.~E.}\ \bibnamefont {Tobar}}, \bibinfo {author}
  {\bibfnamefont {P.}~\bibnamefont {Laurent}}, \bibinfo {author} {\bibfnamefont
  {A.}~\bibnamefont {Clairon}}, \ and\ \bibinfo {author} {\bibfnamefont
  {S.}~\bibnamefont {Bize}},\ }\Doi {10.1103/PhysRevLett.109.080801} {\bibfield
   {journal} {\bibinfo  {journal} {Phys. Rev. Lett.},\ }\textbf {\bibinfo
  {volume} {109}},\ \bibinfo {pages} {080801} (\bibinfo {year}
  {2012})}\BibitemShut {NoStop}%
\bibitem [{\citenamefont {Bondarescu}\ \emph {et~al.}(2012)\citenamefont
  {Bondarescu}, \citenamefont {Bondarescu}, \citenamefont {Hetényi},
  \citenamefont {Boschi}, \citenamefont {Jetzer},\ and\ \citenamefont
  {Balakrishna}}]{Bondarescu01102012}%
  \BibitemOpen
  \bibfield  {author} {\bibinfo {author} {\bibfnamefont {R.}~\bibnamefont
  {Bondarescu}}, \bibinfo {author} {\bibfnamefont {M.}~\bibnamefont
  {Bondarescu}}, \bibinfo {author} {\bibfnamefont {G.}~\bibnamefont
  {Hetényi}}, \bibinfo {author} {\bibfnamefont {L.}~\bibnamefont {Boschi}},
  \bibinfo {author} {\bibfnamefont {P.}~\bibnamefont {Jetzer}}, \ and\ \bibinfo
  {author} {\bibfnamefont {J.}~\bibnamefont {Balakrishna}},\ }\Doi
  {10.1111/j.1365-246X.2012.05636.x} {\bibfield  {journal} {\bibinfo  {journal}
  {Geophysical Journal International},\ }\textbf {\bibinfo {volume} {191}},\
  \bibinfo {pages} {78} (\bibinfo {year} {2012})}\BibitemShut {NoStop}%
\bibitem [{\citenamefont {Drever}\ \emph {et~al.}(1983)\citenamefont {Drever},
  \citenamefont {Hall}, \citenamefont {Kowalski}, \citenamefont {Hough},
  \citenamefont {Ford}, \citenamefont {Munley},\ and\ \citenamefont
  {Ward}}]{drever1983laser}%
  \BibitemOpen
  \bibfield  {author} {\bibinfo {author} {\bibfnamefont {R.}~\bibnamefont
  {Drever}}, \bibinfo {author} {\bibfnamefont {J.~L.}\ \bibnamefont {Hall}},
  \bibinfo {author} {\bibfnamefont {F.}~\bibnamefont {Kowalski}}, \bibinfo
  {author} {\bibfnamefont {J.}~\bibnamefont {Hough}}, \bibinfo {author}
  {\bibfnamefont {G.}~\bibnamefont {Ford}}, \bibinfo {author} {\bibfnamefont
  {A.}~\bibnamefont {Munley}}, \ and\ \bibinfo {author} {\bibfnamefont
  {H.}~\bibnamefont {Ward}},\ }\href@noop {} {\bibfield  {journal} {\bibinfo
  {journal} {Applied Physics B},\ }\textbf {\bibinfo {volume} {31}},\ \bibinfo
  {pages} {97} (\bibinfo {year} {1983})}\BibitemShut {NoStop}%
\bibitem [{\citenamefont {Young}\ \emph {et~al.}(1999)\citenamefont {Young},
  \citenamefont {Cruz}, \citenamefont {Itano},\ and\ \citenamefont
  {Bergquist}}]{young1999visible}%
  \BibitemOpen
  \bibfield  {author} {\bibinfo {author} {\bibfnamefont {B.~C.}\ \bibnamefont
  {Young}}, \bibinfo {author} {\bibfnamefont {F.~C.}\ \bibnamefont {Cruz}},
  \bibinfo {author} {\bibfnamefont {W.~M.}\ \bibnamefont {Itano}}, \ and\
  \bibinfo {author} {\bibfnamefont {J.~C.}\ \bibnamefont {Bergquist}},\ }\Doi
  {10.1103/PhysRevLett.82.3799} {\bibfield  {journal} {\bibinfo  {journal}
  {Phys. Rev. Lett.},\ }\textbf {\bibinfo {volume} {82}},\ \bibinfo {pages}
  {3799} (\bibinfo {year} {1999})}\BibitemShut {NoStop}%
\bibitem [{\citenamefont {Jiang}\ \emph
  {et~al.}(2011){\natexlab{a}}\citenamefont {Jiang}, \citenamefont {Ludlow},
  \citenamefont {Lemke}, \citenamefont {Fox}, \citenamefont {Sherman},
  \citenamefont {Ma},\ and\ \citenamefont {Oates}}]{jiang2011making}%
  \BibitemOpen
  \bibfield  {author} {\bibinfo {author} {\bibfnamefont {Y.}~\bibnamefont
  {Jiang}}, \bibinfo {author} {\bibfnamefont {A.}~\bibnamefont {Ludlow}},
  \bibinfo {author} {\bibfnamefont {N.}~\bibnamefont {Lemke}}, \bibinfo
  {author} {\bibfnamefont {R.}~\bibnamefont {Fox}}, \bibinfo {author}
  {\bibfnamefont {J.}~\bibnamefont {Sherman}}, \bibinfo {author} {\bibfnamefont
  {L.-S.}\ \bibnamefont {Ma}}, \ and\ \bibinfo {author} {\bibfnamefont
  {C.}~\bibnamefont {Oates}},\ }\href@noop {} {\bibfield  {journal} {\bibinfo
  {journal} {Nature Photonics},\ }\textbf {\bibinfo {volume} {5}},\ \bibinfo
  {pages} {158} (\bibinfo {year} {2011}{\natexlab{a}})}\BibitemShut {NoStop}%
\bibitem [{\citenamefont {Numata}\ \emph {et~al.}(2004)\citenamefont {Numata},
  \citenamefont {Kemery},\ and\ \citenamefont {Camp}}]{numata2004thermal}%
  \BibitemOpen
  \bibfield  {author} {\bibinfo {author} {\bibfnamefont {K.}~\bibnamefont
  {Numata}}, \bibinfo {author} {\bibfnamefont {A.}~\bibnamefont {Kemery}}, \
  and\ \bibinfo {author} {\bibfnamefont {J.}~\bibnamefont {Camp}},\ }\href@noop
  {} {\bibfield  {journal} {\bibinfo  {journal} {Physical review letters},\
  }\textbf {\bibinfo {volume} {93}},\ \bibinfo {pages} {250602} (\bibinfo
  {year} {2004})}\BibitemShut {NoStop}%
\bibitem [{\citenamefont {Kessler}\ \emph {et~al.}(2012)\citenamefont
  {Kessler}, \citenamefont {Hagemann}, \citenamefont {Grebing}, \citenamefont
  {Legero}, \citenamefont {Sterr}, \citenamefont {Riehle}, \citenamefont
  {Martin}, \citenamefont {Chen},\ and\ \citenamefont {Ye}}]{Kessler}%
  \BibitemOpen
  \bibfield  {author} {\bibinfo {author} {\bibfnamefont {T.}~\bibnamefont
  {Kessler}}, \bibinfo {author} {\bibfnamefont {C.}~\bibnamefont {Hagemann}},
  \bibinfo {author} {\bibfnamefont {C.}~\bibnamefont {Grebing}}, \bibinfo
  {author} {\bibfnamefont {T.}~\bibnamefont {Legero}}, \bibinfo {author}
  {\bibfnamefont {U.}~\bibnamefont {Sterr}}, \bibinfo {author} {\bibfnamefont
  {F.}~\bibnamefont {Riehle}}, \bibinfo {author} {\bibfnamefont {M.~J.}\
  \bibnamefont {Martin}}, \bibinfo {author} {\bibfnamefont {L.}~\bibnamefont
  {Chen}}, \ and\ \bibinfo {author} {\bibfnamefont {J.}~\bibnamefont {Ye}},\
  }\href {http://dx.doi.org/10.1038/nphoton.2012.217} {\bibfield  {journal}
  {\bibinfo  {journal} {Nat. Photon},\ }\textbf {\bibinfo {volume} {6}},\
  \bibinfo {pages} {687–} (\bibinfo {year} {2012})}\BibitemShut {NoStop}%
\bibitem [{\citenamefont {Martin}\ \emph {et~al.}(2013)\citenamefont {Martin},
  \citenamefont {Bishof}, \citenamefont {Swallows}, \citenamefont {Zhang},
  \citenamefont {Benko}, \citenamefont {von Stecher}, \citenamefont {Gorshkov},
  \citenamefont {Rey},\ and\ \citenamefont {Ye}}]{Martin}%
  \BibitemOpen
  \bibfield  {author} {\bibinfo {author} {\bibfnamefont {M.~J.}\ \bibnamefont
  {Martin}}, \bibinfo {author} {\bibfnamefont {M.}~\bibnamefont {Bishof}},
  \bibinfo {author} {\bibfnamefont {M.~D.}\ \bibnamefont {Swallows}}, \bibinfo
  {author} {\bibfnamefont {X.}~\bibnamefont {Zhang}}, \bibinfo {author}
  {\bibfnamefont {C.}~\bibnamefont {Benko}}, \bibinfo {author} {\bibfnamefont
  {J.}~\bibnamefont {von Stecher}}, \bibinfo {author} {\bibfnamefont {A.~V.}\
  \bibnamefont {Gorshkov}}, \bibinfo {author} {\bibfnamefont {A.~M.}\
  \bibnamefont {Rey}}, \ and\ \bibinfo {author} {\bibfnamefont
  {J.}~\bibnamefont {Ye}},\ }\Doi {10.1126/science.1236929} {\bibfield
  {journal} {\bibinfo  {journal} {Science},\ }\textbf {\bibinfo {volume}
  {341}},\ \bibinfo {pages} {632} (\bibinfo {year} {2013})}\BibitemShut
  {NoStop}%
\bibitem [{\citenamefont {Martin}\ \emph {et~al.}(2011)\citenamefont {Martin},
  \citenamefont {Meiser}, \citenamefont {Thomsen}, \citenamefont {Ye},\ and\
  \citenamefont {Holland}}]{PhysRevA.84.063813}%
  \BibitemOpen
  \bibfield  {author} {\bibinfo {author} {\bibfnamefont {M.~J.}\ \bibnamefont
  {Martin}}, \bibinfo {author} {\bibfnamefont {D.}~\bibnamefont {Meiser}},
  \bibinfo {author} {\bibfnamefont {J.~W.}\ \bibnamefont {Thomsen}}, \bibinfo
  {author} {\bibfnamefont {J.}~\bibnamefont {Ye}}, \ and\ \bibinfo {author}
  {\bibfnamefont {M.~J.}\ \bibnamefont {Holland}},\ }\Doi
  {10.1103/PhysRevA.84.063813} {\bibfield  {journal} {\bibinfo  {journal}
  {Phys. Rev. A},\ }\textbf {\bibinfo {volume} {84}},\ \bibinfo {pages}
  {063813} (\bibinfo {year} {2011})}\BibitemShut {NoStop}%
\bibitem [{\citenamefont {Bonifacio}\ and\ \citenamefont
  {Lugiato}(1978)}]{PhysRevA.18.1129}%
  \BibitemOpen
  \bibfield  {author} {\bibinfo {author} {\bibfnamefont {R.}~\bibnamefont
  {Bonifacio}}\ and\ \bibinfo {author} {\bibfnamefont {L.~A.}\ \bibnamefont
  {Lugiato}},\ }\Doi {10.1103/PhysRevA.18.1129} {\bibfield  {journal} {\bibinfo
   {journal} {Phys. Rev. A},\ }\textbf {\bibinfo {volume} {18}},\ \bibinfo
  {pages} {1129} (\bibinfo {year} {1978})}\BibitemShut {NoStop}%
\bibitem [{\citenamefont {Drummond}\ and\ \citenamefont
  {Walls}(1980)}]{0305-4470-13-2-034}%
  \BibitemOpen
  \bibfield  {author} {\bibinfo {author} {\bibfnamefont {P.~D.}\ \bibnamefont
  {Drummond}}\ and\ \bibinfo {author} {\bibfnamefont {D.~F.}\ \bibnamefont
  {Walls}},\ }\href {http://stacks.iop.org/0305-4470/13/i=2/a=034} {\bibfield
  {journal} {\bibinfo  {journal} {Journal of Physics A: Mathematical and
  General},\ }\textbf {\bibinfo {volume} {13}},\ \bibinfo {pages} {725}
  (\bibinfo {year} {1980})}\BibitemShut {NoStop}%
\bibitem [{\citenamefont {Westergaard}\ \emph {et~al.}(2015)\citenamefont
  {Westergaard}, \citenamefont {Christensen}, \citenamefont {Tieri},
  \citenamefont {Matin}, \citenamefont {Cooper}, \citenamefont {Holland},
  \citenamefont {Ye},\ and\ \citenamefont {Thomsen}}]{JanPRL}%
  \BibitemOpen
  \bibfield  {author} {\bibinfo {author} {\bibfnamefont {P.~G.}\ \bibnamefont
  {Westergaard}}, \bibinfo {author} {\bibfnamefont {B.~T.~R.}\ \bibnamefont
  {Christensen}}, \bibinfo {author} {\bibfnamefont {D.}~\bibnamefont {Tieri}},
  \bibinfo {author} {\bibfnamefont {R.}~\bibnamefont {Matin}}, \bibinfo
  {author} {\bibfnamefont {J.}~\bibnamefont {Cooper}}, \bibinfo {author}
  {\bibfnamefont {M.}~\bibnamefont {Holland}}, \bibinfo {author} {\bibfnamefont
  {J.}~\bibnamefont {Ye}}, \ and\ \bibinfo {author} {\bibfnamefont {J.~W.}\
  \bibnamefont {Thomsen}},\ }\Doi {10.1103/PhysRevLett.114.093002} {\bibfield
  {journal} {\bibinfo  {journal} {Phys. Rev. Lett.},\ }\textbf {\bibinfo
  {volume} {114}},\ \bibinfo {pages} {093002} (\bibinfo {year}
  {2015})}\BibitemShut {NoStop}%
\bibitem [{\citenamefont {Stenholm}\ and\ \citenamefont
  {Lamb~Jr}(1969)}]{Stenholm1969}%
  \BibitemOpen
  \bibfield  {author} {\bibinfo {author} {\bibfnamefont {S.}~\bibnamefont
  {Stenholm}}\ and\ \bibinfo {author} {\bibfnamefont {W.~E.}\ \bibnamefont
  {Lamb~Jr}},\ }\href@noop {} {\bibfield  {journal} {\bibinfo  {journal}
  {Applied Physics Letters},\ }\textbf {\bibinfo {volume} {181}},\ \bibinfo
  {pages} {618} (\bibinfo {year} {1969})}\BibitemShut {NoStop}%
\bibitem [{\citenamefont {Freed}\ and\ \citenamefont {Javan}(1970)}]{Freed}%
  \BibitemOpen
  \bibfield  {author} {\bibinfo {author} {\bibfnamefont {C.}~\bibnamefont
  {Freed}}\ and\ \bibinfo {author} {\bibfnamefont {A.}~\bibnamefont {Javan}},\
  }\href@noop {} {\bibfield  {journal} {\bibinfo  {journal} {Applied Physics
  Letters},\ }\textbf {\bibinfo {volume} {17}},\ \bibinfo {pages} {53}
  (\bibinfo {year} {1970})}\BibitemShut {NoStop}%
\bibitem [{\citenamefont {Stenholm}(2005)}]{Stenholm}%
  \BibitemOpen
  \bibfield  {author} {\bibinfo {author} {\bibfnamefont {S.}~\bibnamefont
  {Stenholm}},\ }\href@noop {} {\emph {\bibinfo {title} {Foundations of Laser
  Spectroscopy}}}\ (\bibinfo  {publisher} {Dover Publications},\ \bibinfo
  {year} {2005})\BibitemShut {NoStop}%
\bibitem [{\citenamefont {Levenson}(2012)}]{levenson2012introduction}%
  \BibitemOpen
  \bibfield  {author} {\bibinfo {author} {\bibfnamefont {M.}~\bibnamefont
  {Levenson}},\ }\href@noop {} {\emph {\bibinfo {title} {Introduction to
  Nonlinear Laser Spectroscopy}}}\ (\bibinfo  {publisher} {Elsevier},\ \bibinfo
  {year} {2012})\BibitemShut {NoStop}%
\bibitem [{\citenamefont {Agarwal}\ and\ \citenamefont
  {Zhu}(1992)}]{PhysRevA.46.479}%
  \BibitemOpen
  \bibfield  {author} {\bibinfo {author} {\bibfnamefont {G.~S.}\ \bibnamefont
  {Agarwal}}\ and\ \bibinfo {author} {\bibfnamefont {Y.}~\bibnamefont {Zhu}},\
  }\Doi {10.1103/PhysRevA.46.479} {\bibfield  {journal} {\bibinfo  {journal}
  {Phys. Rev. A},\ }\textbf {\bibinfo {volume} {46}},\ \bibinfo {pages} {479}
  (\bibinfo {year} {1992})}\BibitemShut {NoStop}%
\bibitem [{\citenamefont {Tallet}(1994)}]{Tallet:94}%
  \BibitemOpen
  \bibfield  {author} {\bibinfo {author} {\bibfnamefont {A.}~\bibnamefont
  {Tallet}},\ }\Doi {10.1364/JOSAB.11.001336} {\bibfield  {journal} {\bibinfo
  {journal} {J. Opt. Soc. Am. B},\ }\textbf {\bibinfo {volume} {11}},\ \bibinfo
  {pages} {1336} (\bibinfo {year} {1994})}\BibitemShut {NoStop}%
\bibitem [{\citenamefont {Gripp}\ \emph {et~al.}(1997)\citenamefont {Gripp},
  \citenamefont {Mielke},\ and\ \citenamefont {Orozco}}]{PhysRevA.56.3262}%
  \BibitemOpen
  \bibfield  {author} {\bibinfo {author} {\bibfnamefont {J.}~\bibnamefont
  {Gripp}}, \bibinfo {author} {\bibfnamefont {S.~L.}\ \bibnamefont {Mielke}}, \
  and\ \bibinfo {author} {\bibfnamefont {L.~A.}\ \bibnamefont {Orozco}},\ }\Doi
  {10.1103/PhysRevA.56.3262} {\bibfield  {journal} {\bibinfo  {journal} {Phys.
  Rev. A},\ }\textbf {\bibinfo {volume} {56}},\ \bibinfo {pages} {3262}
  (\bibinfo {year} {1997})}\BibitemShut {NoStop}%
\bibitem [{\citenamefont {Ficek}\ and\ \citenamefont
  {Freedhoff}(1993)}]{PhysRevA.48.3092}%
  \BibitemOpen
  \bibfield  {author} {\bibinfo {author} {\bibfnamefont {Z.}~\bibnamefont
  {Ficek}}\ and\ \bibinfo {author} {\bibfnamefont {H.~S.}\ \bibnamefont
  {Freedhoff}},\ }\Doi {10.1103/PhysRevA.48.3092} {\bibfield  {journal}
  {\bibinfo  {journal} {Phys. Rev. A},\ }\textbf {\bibinfo {volume} {48}},\
  \bibinfo {pages} {3092} (\bibinfo {year} {1993})}\BibitemShut {NoStop}%
\bibitem [{\citenamefont {Agarwal}\ \emph {et~al.}(1991)\citenamefont
  {Agarwal}, \citenamefont {Zhu}, \citenamefont {Gauthier},\ and\ \citenamefont
  {Mossberg}}]{Agarwal:91}%
  \BibitemOpen
  \bibfield  {author} {\bibinfo {author} {\bibfnamefont {G.~S.}\ \bibnamefont
  {Agarwal}}, \bibinfo {author} {\bibfnamefont {Y.}~\bibnamefont {Zhu}},
  \bibinfo {author} {\bibfnamefont {D.~J.}\ \bibnamefont {Gauthier}}, \ and\
  \bibinfo {author} {\bibfnamefont {T.~W.}\ \bibnamefont {Mossberg}},\ }\Doi
  {10.1364/JOSAB.8.001163} {\bibfield  {journal} {\bibinfo  {journal} {J. Opt.
  Soc. Am. B},\ }\textbf {\bibinfo {volume} {8}},\ \bibinfo {pages} {1163}
  (\bibinfo {year} {1991})}\BibitemShut {NoStop}%
\bibitem [{\citenamefont {Press}\ \emph {et~al.}(2007)\citenamefont {Press},
  \citenamefont {Teukolsky}, \citenamefont {Vetterling},\ and\ \citenamefont
  {Flannery}}]{ASC}%
  \BibitemOpen
  \bibfield  {author} {\bibinfo {author} {\bibfnamefont {W.~H.}\ \bibnamefont
  {Press}}, \bibinfo {author} {\bibfnamefont {S.~A.}\ \bibnamefont
  {Teukolsky}}, \bibinfo {author} {\bibfnamefont {W.~T.}\ \bibnamefont
  {Vetterling}}, \ and\ \bibinfo {author} {\bibfnamefont {B.~P.}\ \bibnamefont
  {Flannery}},\ }\href@noop {} {\emph {\bibinfo {title} {Numerical Recipes 3rd
  Edition: The Art of Scientific Computing}}},\ \bibinfo {edition} {3rd}\ ed.\
  (\bibinfo  {publisher} {Cambridge University Press},\ \bibinfo {address} {New
  York, NY, USA},\ \bibinfo {year} {2007})\ ISBN \bibinfo {isbn} {0521880688,
  9780521880688}\BibitemShut {NoStop}%
\bibitem [{\citenamefont {Riedmann}\ \emph {et~al.}(2012)\citenamefont
  {Riedmann}, \citenamefont {Kelkar}, \citenamefont {W{\"u}bbena},
  \citenamefont {Pape}, \citenamefont {Kulosa}, \citenamefont {Zipfel},
  \citenamefont {Fim}, \citenamefont {R{\"u}hmann}, \citenamefont {Friebe},
  \citenamefont {Ertmer},\ and\ \citenamefont {Rasel}}]{MgPaper}%
  \BibitemOpen
  \bibfield  {author} {\bibinfo {author} {\bibfnamefont {M.}~\bibnamefont
  {Riedmann}}, \bibinfo {author} {\bibfnamefont {H.}~\bibnamefont {Kelkar}},
  \bibinfo {author} {\bibfnamefont {T.}~\bibnamefont {W{\"u}bbena}}, \bibinfo
  {author} {\bibfnamefont {A.}~\bibnamefont {Pape}}, \bibinfo {author}
  {\bibfnamefont {A.}~\bibnamefont {Kulosa}}, \bibinfo {author} {\bibfnamefont
  {K.}~\bibnamefont {Zipfel}}, \bibinfo {author} {\bibfnamefont
  {D.}~\bibnamefont {Fim}}, \bibinfo {author} {\bibfnamefont {S.}~\bibnamefont
  {R{\"u}hmann}}, \bibinfo {author} {\bibfnamefont {J.}~\bibnamefont {Friebe}},
  \bibinfo {author} {\bibfnamefont {W.}~\bibnamefont {Ertmer}}, \ and\ \bibinfo
  {author} {\bibfnamefont {E.}~\bibnamefont {Rasel}},\ }\Doi
  {10.1103/PhysRevA.86.043416} {\bibfield  {journal} {\bibinfo  {journal}
  {Phys. Rev. A},\ }\textbf {\bibinfo {volume} {86}},\ \bibinfo {pages}
  {043416} (\bibinfo {year} {2012})}\BibitemShut {NoStop}%
\bibitem [{\citenamefont {Kohno}\ \emph {et~al.}(2009)\citenamefont {Kohno},
  \citenamefont {Yasuda}, \citenamefont {Hosaka}, \citenamefont {Inaba},
  \citenamefont {Nakajima},\ and\ \citenamefont {Hong}}]{YbPaper}%
  \BibitemOpen
  \bibfield  {author} {\bibinfo {author} {\bibfnamefont {T.}~\bibnamefont
  {Kohno}}, \bibinfo {author} {\bibfnamefont {M.}~\bibnamefont {Yasuda}},
  \bibinfo {author} {\bibfnamefont {K.}~\bibnamefont {Hosaka}}, \bibinfo
  {author} {\bibfnamefont {H.}~\bibnamefont {Inaba}}, \bibinfo {author}
  {\bibfnamefont {Y.}~\bibnamefont {Nakajima}}, \ and\ \bibinfo {author}
  {\bibfnamefont {F.-L.}\ \bibnamefont {Hong}},\ }\href
  {http://stacks.iop.org/1882-0786/2/i=7/a=072501} {\bibfield  {journal}
  {\bibinfo  {journal} {Applied Physics Express},\ }\textbf {\bibinfo {volume}
  {2}},\ \bibinfo {pages} {072501} (\bibinfo {year} {2009})}\BibitemShut
  {NoStop}%
\bibitem [{\citenamefont {Dammalapati}\ \emph {et~al.}(2009)\citenamefont
  {Dammalapati}, \citenamefont {Norris}, \citenamefont {Maguire}, \citenamefont
  {Borkowski},\ and\ \citenamefont {Riis}}]{CaPaper}%
  \BibitemOpen
  \bibfield  {author} {\bibinfo {author} {\bibfnamefont {U.}~\bibnamefont
  {Dammalapati}}, \bibinfo {author} {\bibfnamefont {I.}~\bibnamefont {Norris}},
  \bibinfo {author} {\bibfnamefont {L.}~\bibnamefont {Maguire}}, \bibinfo
  {author} {\bibfnamefont {M.}~\bibnamefont {Borkowski}}, \ and\ \bibinfo
  {author} {\bibfnamefont {E.}~\bibnamefont {Riis}},\ }\href
  {http://stacks.iop.org/0957-0233/20/i=9/a=095303} {\bibfield  {journal}
  {\bibinfo  {journal} {Measurement Science and Technology},\ }\textbf
  {\bibinfo {volume} {20}},\ \bibinfo {pages} {095303} (\bibinfo {year}
  {2009})}\BibitemShut {NoStop}%
\bibitem [{\citenamefont {{Rakhmanov}}\ \emph {et~al.}(2002)\citenamefont
  {{Rakhmanov}}, \citenamefont {{Savage}}, \citenamefont {{Reitze}},\ and\
  \citenamefont {{Tanner}}}]{2002PhLA..305..239R}%
  \BibitemOpen
  \bibfield  {author} {\bibinfo {author} {\bibfnamefont {M.}~\bibnamefont
  {{Rakhmanov}}}, \bibinfo {author} {\bibfnamefont {R.~L.}\ \bibnamefont
  {{Savage}}}, \bibinfo {author} {\bibfnamefont {D.~H.}\ \bibnamefont
  {{Reitze}}}, \ and\ \bibinfo {author} {\bibfnamefont {D.~B.}\ \bibnamefont
  {{Tanner}}},\ }\Doi {10.1016/S0375-9601(02)01469-X} {\bibfield  {journal}
  {\bibinfo  {journal} {Physics Letters A},\ }\textbf {\bibinfo {volume}
  {305}},\ \bibinfo {pages} {239} (\bibinfo {year} {2002})},\ \Eprint
  {http://arxiv.org/abs/physics/0110061} {physics/0110061} \BibitemShut
  {NoStop}%
\bibitem [{\citenamefont {Dick}(1987)}]{Dick}%
  \BibitemOpen
  \bibfield  {author} {\bibinfo {author} {\bibfnamefont {G.~J.}\ \bibnamefont
  {Dick}},\ }\href@noop {} {\bibfield  {journal} {\bibinfo  {journal} {Proc.
  Precise Time and Time Interval Meeting},\ \bibinfo {pages} {133}} (\bibinfo
  {year} {1987})}\BibitemShut {NoStop}%
\bibitem [{\citenamefont {Westergaard}\ \emph {et~al.}(2010)\citenamefont
  {Westergaard}, \citenamefont {Lodewyck},\ and\ \citenamefont
  {Lemonde}}]{westergaard2010minimizing}%
  \BibitemOpen
  \bibfield  {author} {\bibinfo {author} {\bibfnamefont {P.}~\bibnamefont
  {Westergaard}}, \bibinfo {author} {\bibfnamefont {J.}~\bibnamefont
  {Lodewyck}}, \ and\ \bibinfo {author} {\bibfnamefont {P.}~\bibnamefont
  {Lemonde}},\ }\href@noop {} {\bibfield  {journal} {\bibinfo  {journal}
  {Ultrasonics, Ferroelectrics, and Frequency Control, IEEE Transactions on},\
  }\textbf {\bibinfo {volume} {57}},\ \bibinfo {pages} {623} (\bibinfo {year}
  {2010})}\BibitemShut {NoStop}%
\bibitem [{\citenamefont {Jiang}\ \emph
  {et~al.}(2011){\natexlab{b}}\citenamefont {Jiang}, \citenamefont {Ludlow},
  \citenamefont {Lemke}, \citenamefont {Sherman}, \citenamefont {Von~Stecher},
  \citenamefont {Fox}, \citenamefont {Ma}, \citenamefont {Rey},\ and\
  \citenamefont {Oates}}]{5977827}%
  \BibitemOpen
  \bibfield  {author} {\bibinfo {author} {\bibfnamefont {Y.}~\bibnamefont
  {Jiang}}, \bibinfo {author} {\bibfnamefont {A.}~\bibnamefont {Ludlow}},
  \bibinfo {author} {\bibfnamefont {N.}~\bibnamefont {Lemke}}, \bibinfo
  {author} {\bibfnamefont {J.}~\bibnamefont {Sherman}}, \bibinfo {author}
  {\bibfnamefont {J.}~\bibnamefont {Von~Stecher}}, \bibinfo {author}
  {\bibfnamefont {R.}~\bibnamefont {Fox}}, \bibinfo {author} {\bibfnamefont
  {L.-S.}\ \bibnamefont {Ma}}, \bibinfo {author} {\bibfnamefont
  {A.}~\bibnamefont {Rey}}, \ and\ \bibinfo {author} {\bibfnamefont
  {C.}~\bibnamefont {Oates}},\ }in\ \Doi {10.1109/FCS.2011.5977827} {\emph
  {\bibinfo {booktitle} {Frequency Control and the European Frequency and Time
  Forum (FCS), 2011 Joint Conference of the IEEE International}}}\ (\bibinfo
  {year} {2011})\ pp.\ \bibinfo {pages} {1--3},\ ISSN \bibinfo {issn}
  {1075-6787}\BibitemShut {NoStop}%
\bibitem [{\citenamefont {Yang}\ \emph {et~al.}(2015)\citenamefont {Yang},
  \citenamefont {Pandey}, \citenamefont {Pramod}, \citenamefont {Leroux},
  \citenamefont {Kwong}, \citenamefont {Hajiyev}, \citenamefont {Chia},
  \citenamefont {Fang},\ and\ \citenamefont {Wilkowski}}]{AtomFlux}%
  \BibitemOpen
  \bibfield  {author} {\bibinfo {author} {\bibfnamefont {T.}~\bibnamefont
  {Yang}}, \bibinfo {author} {\bibfnamefont {K.}~\bibnamefont {Pandey}},
  \bibinfo {author} {\bibfnamefont {M.~S.}\ \bibnamefont {Pramod}}, \bibinfo
  {author} {\bibfnamefont {F.}~\bibnamefont {Leroux}}, \bibinfo {author}
  {\bibfnamefont {C.}~\bibnamefont {Kwong}}, \bibinfo {author} {\bibfnamefont
  {E.}~\bibnamefont {Hajiyev}}, \bibinfo {author} {\bibfnamefont
  {Z.}~\bibnamefont {Chia}}, \bibinfo {author} {\bibfnamefont {B.}~\bibnamefont
  {Fang}}, \ and\ \bibinfo {author} {\bibfnamefont {D.}~\bibnamefont
  {Wilkowski}},\ }\href@noop {} { (\bibinfo {year} {2015})},\ \Eprint
  {http://arxiv.org/abs/0902.0885} {{arXiv}:0902.0885} \BibitemShut {NoStop}%
\end{thebibliography}%

\end{document}